\documentclass[12pt]{article}

\usepackage{a4wide}
\usepackage{amssymb}
\usepackage{amsmath}
\usepackage{bm}

\usepackage{scrextend}
\addtokomafont{labelinglabel}{\sffamily}

\usepackage{hyperref}
 \numberwithin{equation}{section}

\usepackage{amsfonts}

\usepackage{epsfig}

\newcommand{\be}{\begin{equation}}
\newcommand{\ee}{\end{equation}}
\newcommand{\bea}{\begin{eqnarray}}
\newcommand{\eea}{\end{eqnarray}}

\def\tr{{\rm tr}}
\def\bfphi{\mbox{\boldmath $\phi$}}
\begin{document}

\setcounter{table}{0}

\begin{flushright}\footnotesize

\texttt{ICCUB-18-020}

\end{flushright}

\mbox{}
\vspace{0truecm}
\linespread{1.1}

\vspace{0.5truecm}

\centerline{\Large \bf Late-time Cosmic Acceleration from Compactification}

\vspace{1.3truecm}

\centerline{
    {\large \bf J. G. Russo${}^{a,b}$}
   {\bf and} 
   {\large \bf P. K. Townsend${}^{c}$}
 }

\vspace{0.8cm}

\begin{labeling}{u}

\item [${}^a$]{{\it Instituci\'o Catalana de Recerca i Estudis Avan\c{c}ats (ICREA),\\
Pg. Lluis Companys, 23, 08010 Barcelona, Spain.}}

\item [${}^b$]{{\it  Departament de F\' \i sica Cu\' antica i Astrof\'\i sica and Institut de Ci\`encies del Cosmos,\\ 
Universitat de Barcelona, Mart\'i Franqu\`es, 1, 08028
Barcelona, Spain. }}

\item [${}^c$]
{{\it Department of Applied
Mathematics and Theoretical Physics,\\ 
Centre for Mathematical
Sciences, University of Cambridge,\\
Wilberforce Road, Cambridge, CB3
0WA, UK.  }} 

\end{labeling}

\noindent {\it E-Mail:}  {\texttt jorge.russo@icrea.cat, p.k.townsend@damtp.cam.ac.uk} 

\vspace{1.2cm}

\centerline{\bf ABSTRACT}
\medskip
 
 We investigate the implications of energy conditions on cosmological compactification solutions of the higher-dimensional
Einstein field equations. It is known that the  Strong Energy Condition forbids time-independent compactifications to de Sitter space
but allows  time-dependent compactifications to  other (homogeneous and isotropic) expanding universes  that undergo a
{\it transient} period of acceleration.  Here we show that the same assumptions allow compactification to FLRW universes 
undergoing {late-time} accelerated expansion; the late-time stress tensor is a perfect fluid  but with a lower bound on the 
pressure/energy-density ratio that excludes de Sitter but allows accelerated power-law expansion. The compact space 
undergoes a  decelerating  expansion that leads to decompactification, but on an arbitrarily long timescale.

\noindent

\newpage

%%%%%%%%%%%%%%%%%%%%%%%%%%%%%%%%%%%%%%%%
\tableofcontents

%%%%%%%%%%%%%%%%%%%%%%%%%%%
\section{Introduction}
\setcounter{equation}{0}

Observations indicating that the expansion of the Universe is accelerating are compatible with 
General Relativity (GR)  if one postulates a dark energy contribution to the matter stress tensor. Equivalently, one may 
modify Einstein's field equations to include a positive cosmological constant, as originally proposed on other grounds by 
Einstein himself. Either way, the predicted late-time geometry of an expanding universe is that of de Sitter (dS) but 
difficulties facing any attempt to formulate a consistent quantum theory of gravity in de Sitter space \cite{Witten:2001kn}  suggest that 
current observations may have some other explanation that does not lead to the prediction of a late-time dS universe.

In the context of scalar-tensor theories of gravity in which a dS vacuum corresponds to a  minimum value $V_0>0$ of a scalar-field potential 
$V$, one can propose a function $V$ that has no such minima but does have regions in field space of positive $V$ with $|\nabla V|$ sufficiently
small that accelerated expansion is both possible and sufficiently slowly varying to be compatible with current observations. 
Scalar fields that make this possible have been dubbed ``quintessence''. However, some of the difficulties facing the formulation 
of a consistent quantum theory of gravity in de Sitter space, in particular the future cosmological event horizon associated with late-time 
accelerated expansion,  are still present  in quintessence models \cite{Hellerman:2001yi,Fischler:2001yj}. 

It is natural to ask what string/M-theory has to say about dS vacua and quintessence. In particular, 
one may ask whether the 10/11-dimensional  supergravity theories that arise as low-energy effective 
field theories for string/M-theory  permit dS vacua.  A partial answer to this question is provided by a no-go theorem of 
Gibbons \cite{Gibbons:1984kp} which states that the $D$-dimensional Einstein field equations do not permit a time-independent 
compactification to a dS spacetime of dimension $d<D$ if the stress tensor satisfies the Strong Energy Condition (SEC);  
this follows from a simple inequality implied by the higher-dimensional SEC. Although the SEC is not as fundamental as the Dominant Energy Condition (DEC), 
it is satisfied (with one exception to be mentioned below) by D=10/11  supergravity  theories, so this rules out time-independent compactifications to dS of these theories. 
 This no-go theorem was rediscovered in a  string/M-theory context by Maldacena and Nu\~nez \cite{Maldacena:2000mw}, who extended the result to 
the massive IIA $D=10$ supergravity; this has a dilaton potential that violates the SEC, but a lower-dimensional dS universe is still unobtainable 
by a time-independent compactification.

In principle, this Gibbons-Maldacena-Nu\~nez (GMN) no-go theorem  might be circumvented by relaxing the usual  non-singularity condition on the compact spaces used
for compactification, since some geometrical singularities are innocuous in String/M-theory. An influential model of this type
(incorporating $D$-branes and, crucially, orientifold planes)  has been argued to yield a dS vacuum \cite{Kachru:2003aw} (see \cite{Cicoli:2018kdo} for a recent overview). 
However, despite more than a decade of active investigation  on this topic it was possible to ask less than one year ago
``What if string theory has no de Sitter vacua?''\cite{Danielsson:2018ztv}.  More recently, it has been suggested  that dS vacua belong to the string theory ``swampland''\cite{Obied:2018sgi}.   

Another potential way around the GMN no-go theorem is to relax the condition of  time-independence.
If this is done then   it is important to distinguish between $d$-metrics in different conformal frames; these differ by a scalar-field dependent conformal 
factor which may become time-dependent for time-dependent compactifications. It is usual to insist on  Einstein conformal frame because the $d$-dimensional 
Newton constant is otherwise  time-dependent, and this would be in conflict with observations for $d=4$.  The first attempt to generalize
the GMN theorem to allow for a time-dependent compact space yielded the conclusion that the SEC still prevents compactification to dS 
 \cite{Teo:2004hq}, and some analogous results were presented  more recently in \cite{Obied:2018sgi}.  This conclusion, however, is based 
on some unnecessary assumptions that we make explicit here. 

Our conclusion is {\it not} that compactification to dS is still excluded when time-dependence is allowed, 
but rather that any example of it must implement the Einstein-frame condition in a way that differs from the way it has been implemented previously, 
at least in those time-dependendent cosmological compactifications of which we are aware.  It is not clear to us how this loophole can be exploited in practice, so we focus 
here on a different question: assuming that the Einstein-frame condition is implemented in the customary way, and that compactification to dS 
is then excluded (as we confirm) is {\it late-time} cosmic acceleration also excluded? 

 It is known that the higher-dimensional SEC does not prevent compactification to 
an Einstein-frame FLRW universe that undergoes accelerated expansion for some finite period  \cite{Townsend:2003fx}.  For a  $D$-dimensional stress tensor 
satisfying the SEC, many other time-dependent  compactifications have been found  that lead to a similar transient cosmic acceleration of the lower-dimensional 
Einstein-frame metric  \cite{Cornalba:2002fi,Ohta:2003pu,Roy:2003nd,Wohlfarth:2003ni,Chen:2003ij,Gutperle:2003kc} but none with {\it late-time} acceleration. 
Indeed, this appears to be ruled out by Teo's  generalized no-go theorem  \cite{Teo:2004hq} but the proof implicitly uses the {\it vacuum} Einstein equations in the higher 
dimension. This  leaves open the possibility that some compactification to an FLRW universe undergoing late-time acceleration might be achieved by a non-zero
stress tensor satisfying the SEC in the higher dimension.  

An FLRW cosmology that undergoes late-time accelerated expansion can be expected to approach a scaling solution of the 
Einstein equations for which the accelerated expansion is eternal because these solutions are late-time attractors 
(in contrast to non-accelerating scaling solutions).  If the SEC were to forbid compactification to eternally accelerating universes then this would constitute strong evidence against 
compactification to any universe that undergoes late-time acceleration (i.e. with an expansion rate that is strictly increasing 
from some  time $t_0$). However, what we find is that the SEC does {\it not} forbid compactification to eternally accelerating 
universes. Specifically, we exhibit cosmological compactifications of the $D$-dimensional Einstein equations, with a stress tensor satisfying 
both the SEC and the DEC, such that the lower dimensional FLRW universe undergoes a power-law accelerated expansion. 
A field theory realization of the $D$-dimensional stress tensor, which could allow the  exploration of more general solutions, remains an open 
problem, as does any connection to string/M-theory. 

An interesting feature of these  cosmological compactifications  is that the compact space is also expanding,  which implies an ultimate
decompactification. However, it is remarkable that an accelerating expansion of the  FLRW universe requires a {\it decelerating}
expansion of the compact space.  We explore the implications for our Universe on the assumption that it arises from some compactification 
of this type.  Our conclusion is that the decompactification time (which we define as the  time at which effects of Kaluza-Klein particles would show up in current accelerator experiments) 
is arbitrarily long, and easily much longer than the current age of the universe.  

From a $d$-dimensional perspective,  the conclusion of the GMN no-go theorem can be restated as the condition
that the scalar potential $V$ of the effective $d$-dimensional theory cannot have stationary points with $V=V_0 >0$ 
\cite{Townsend:2001ea}. The occurrence of  transient cosmic acceleration is  then simply explained in terms of the properties 
of such potentials \cite{Emparan:2003gg}:  given a  potential that is positive in some  region of field space,  there exist initial conditions such that  this 
region is entered,  at which point the motion in field space is  `uphill'. If there are no stationary points
 in the region of positive potential then this uphill motion must end, at which moment the  FLRW universe will be 
 accelerating. How long it continues to accelerate depends on details of the potential, in particular its gradient $\nabla V$. Late-time acceleration 
 will occur only if the path in field space  is such that $V$ remains positive and the late-time value of $|\nabla V|/V$ is subcritical.  
 
In our examples of late-time accelerated expansion, the $d$-dimensional FLRW universe is filled with a perfect fluid with 
 a constant pressure/energy-density ratio $w$, but this constant is restricted to lie in a particular interval.
 In the four-dimensional case, 
  \begin{equation}\label{newng}
 -\frac{1}{2}< w <  - \frac{1}{3} \ \qquad \quad (d=4). 
 \end{equation}
 The upper bound on $w$ is simply a consequence of our assumption of accelerated expansion; the lower bound is  implied by the 
 higher-dimensional SEC.   Since $w=-1$ corresponds to a dS universe, this result constitutes a significant generalization of  the GMN no-go theorem 
 to the time-dependent case, on assumptions that are the same as (or weaker than) those used previously 
 (e.g.  \cite{Obied:2018sgi,Teo:2004hq})
 to rule out time-dependent compactifications to dS.  Of course, the restrictions on $w$ apply only at late-times; current observations are compatible with 
 $w=-1$ but they are also compatible with a time-dependent increasing $w$ \cite{Aghanim:2018eyx}. 
  
A feature of FLRW universes undergoing power-law expansion is that they can also be realized as solutions of the $d$-dimensional Einstein equations 
for gravity coupled to a scalar field $\phi$ with positive `exponential' potential $V\propto e^{\lambda\phi}$, for constant $\lambda =V'/V$ \cite{Ferreira:1997hj} 
(see \cite{Townsend:2001ea} for the $d$-dimensional generalization).  In fact, the general  FLRW solution of such a model can be found exactly \cite{Russo:2004ym}; it approaches
 a power-law solution at early and late times, with late-time acceleration only if  $\lambda<\lambda_c$, where $\lambda_c$ is a `critical' value of 
 order unity (its precise value is both $d$-dependent and convention dependent). 
 
 Scalar-tensor theories of this type typically arise as consistent truncations 
 of compactified D=10/11 supergravity theories and it was conjectured in \cite{Townsend:2003qv} (based on a variety of examples) that this always 
 leads to  $\lambda>\lambda_c$; this conjecture was proved in \cite{Teo:2004hq} but subject to a restrictive implicit assumption (that
 we make explicit here). A subsequent study for generic multi-scalar theories with positive potential $V(\bfphi)$ focused on the time evolution of the 
 vector of ``characteristic functions''  $\partial \ln V/\partial \bfphi$ \cite{Townsend:2004zp}.  These results are  presumably  related to the more recent 
 conjecture that positive  potentials arising from string/M-theory  compactifications are such that  $|\nabla V|/V \ge c$ for some  constant  
 $c$ \cite{Obied:2018sgi}; we  comment on this proposal in our conclusions.  
 
The organization is as follows. We first derive the {\it general } SEC inequality for time-dependent  compactifications to an FLRW universe. 
This is a new result since previous SEC inequalities have all made assumptions that are not required by FLRW isometries. 
We discuss the difficulty of drawing definite conclusions from this inequality, and then proceed to make the  simplifying assumption concerning implementation of the Einstein-frame condition that is 
implicit in previous investigations of time-dependent compactifications. This takes us to a simpler SEC inequality; we
analyse its implications, confirming that it excludes compactification to dS. We then analyse the implications for compactifications 
to FLRW universes with power-law expansion, finding examples for which the expansion is accelerating. In section 3 we turn to an
investigation of these examples,  examining the implications for decompactification, showing that  the required stress tensor in $D$-dimensions satisfies the 
DEC in addition to the SEC,  and  finding the restriction  (\ref{newng}) on the lower-dimensional stress tensor. As the compact space in our examples
is a scale factor times a Ricci-flat metric, we briefly consider the implications of relaxing the Ricci flat condition.  We conclude with a summary of our main results and 
a further discussion of constraints on scalar potentials in the $d$-dimensional  effective scalar-tensor gravity theory.

%%%%%%%%%%%%%%%%%%%%%%%%%%%%%%%%%%%%%%%%%%
%%%%%%%%%%%%%%%%%%%%%%%%%%%%%%%%%%%%%%%%%%%

 \section{Warped compactifications to FLRW cosmologies}

The general $d$-dimensional FLRW spacetime has a metric of the form
\begin{equation}\label{FLRWmetric}
ds^2_{FLRW} \equiv  g_{\mu\nu} dx^\mu dx^\nu  = -dt^2 + S^2(t)\,  \bar g_{ij} dx^i dx^j\, ,  
\end{equation}
where $S(t)$ is the scale factor as a function of standard FLRW time, and $\bar g_{ij}$ is the metric in local coordinates
$\{x^i; \ i=1,\cdots, d-1\}$ for a maximally-symmetric $(d-1)$-space with constant  curvature $k$, which we leave un-normalized.  
Our starting point will be a $D$-dimensional manifold that is  topologically a product of 
this FLRW spacetime with a compact $n$-dimensional manifold $B$ (so $D=d+n$). The most general $D$-metric compatible 
with the FLRW isometries, which are those of $\bar g_{ij}$,  has the form
\begin{equation}
ds^2_D =  \Omega^2(y;t) ds^2_{FLRW} + h_{\alpha\beta}(y;t) dy^\alpha dy^\beta\, , 
\end{equation}
where $h_{\alpha\beta}$ is the metric on $B$ in local coordinates $\{y^\alpha; \alpha=1,\dots,n\}$, and  $\Omega$ is a nowhere-zero
`warp factor': a scalar function on $B$.   Notice that the FLRW isometries permit not only the metric on the 
compact manifold $B$ to be time-dependent but also the warp factor. 

The  condition for $ds^2_{FLRW}$  to be an Einstein frame metric for the effective $d$-dimensional gravity theory is
\begin{equation}\label{Eframe}
\int_B d^ny\,  \sqrt{\det h}\, \Omega^{d-2}  = G_D/G_d\, , 
\end{equation}
where the constant on the right-hand side is the ratio of the Newton constants in the higher and lower spacetime dimensions. 
This motivates the following notation: for any scalar $\Phi$ on $B$, we shall define
\begin{equation}
\langle \Phi\rangle  = (G_d/G_D)\int_B d^ny\,  \sqrt{\det h}\, \Omega^{d-2} \Phi\, . 
\end{equation}
The Einstein-frame condition in this notation is $\langle 1\rangle =1$, which means that $\langle\Phi\rangle$ can be
interpreted as an average of $\Phi$ over $B$; this means, in particular,  that $\langle\Phi\rangle=\Phi$ when 
$\Phi$ is independent of position on $B$.

As the Einstein-frame condition must hold for all $t$ we may deduce, by taking its time
derivative, the following ``first-order'' Einstein-frame condition:
\begin{equation}\label{EFCderived}
\frac12 \langle \tr \left(h^{-1} \dot h\right)\rangle = -(d-2)\langle \dot\Omega/\Omega \rangle\, . 
\end{equation}
By taking another derivative we deduce the ``second-order'' Einstein-frame condition
  \begin{equation}\label{ddoth}
 \frac12 \langle \tr \left(h^{-1}\ddot h\right) \rangle = \frac12 \langle \tr \left(h^{-1}\dot h\right)^2\rangle  
 + (d-2) \langle \left[(\dot\Omega/\Omega)^2 - (\ddot\Omega/\Omega)\right] \rangle -  \langle X^2\rangle\, ,  
 \end{equation}
 where 
 \begin{equation}
     X\equiv \frac12 \tr \left(h^{-1} \dot h\right)  + (d-2)\dot\Omega/\Omega\, . 
%\nonumber
 \end{equation}
 Notice that $X$ averages to zero as a consequence of (\ref{EFCderived}); however, this does not imply that 
 the same is true of $X^2$. 
 
 %%%%%%%%%%%%%%%%%%%%%%
\subsection{The Strong Energy Condition}
 
 The SEC on the stress tensor in $D$ dimensions implies, via the Einstein field equations, that the time-time component of the Ricci tensor is non-negative. 
 A direct calculation of this Ricci tensor component yields
\begin{eqnarray}\label{r00}
R_{00} &=& -\frac12 \tr\left(h^{-1}\ddot h\right) -(d-1)\left( \frac{\ddot S}{S} + \frac{\ddot\Omega}{\Omega}\right) + \frac{1}{d}\Omega^{2-d} \nabla^2\Omega^d  \\
&& +\ (d-1)\left[(\dot\Omega/\Omega)^2 - (\dot\Omega/\Omega)(\dot S/S)\right] + \frac12\left(\frac{\dot\Omega}{\Omega}\right)\tr\left(h^{-1}\dot h\right) +
 \frac14 \tr\left[\left(h^{-1}\dot h\right)^2\right] \, . \nonumber
\end{eqnarray}
If we average over $B$ then we may use the second-order Einstein-frame condition (\ref{ddoth}),  together with 
\begin{equation}
\left\langle \Omega^{2-d} \nabla^2 \Omega^d \right\rangle =: \int_B d^ny \sqrt{\det h} \nabla^2 \Omega 
= \int_B d^ny \nabla_\alpha \left(\sqrt{\det h} \, h^{\alpha\beta} \partial_\beta\Omega^d\right) = 0\, , 
\end{equation}
to deduce that 
\begin{eqnarray}
 \left\langle R_{00} \right\rangle &=&  - (d-1)\left[ (\ddot S/S) + (\dot S/S) \left\langle \dot\Omega/\Omega\right\rangle  \right]
 - \left\langle \ddot\Omega/\Omega\right\rangle - (d-3) \left\langle (\dot\Omega/\Omega)^2\right\rangle \nonumber \\
 && - \frac14 \langle \tr (h^{-1}\dot h)^2 \rangle + 
 \left\langle X\left[X+ \dot\Omega/\Omega\right] \right\rangle \, . 
  \end{eqnarray}
 The left hand side is non negative if the SEC is satisfied, so a {\it necessary} condition for the SEC
 to hold is that the right hand side is non-negative. This inequality is 
 \begin{eqnarray}\label{halfway}
 && - (d-1)\left[ (\ddot S/S) + (\dot S/S) \left\langle \dot\Omega/\Omega\right\rangle  \right]
 - \left\langle \ddot\Omega/\Omega\right\rangle - (d-3) \left\langle (\dot\Omega/\Omega)^2\right\rangle 
 + \left\langle X\left[X+ \dot\Omega/\Omega\right]\right\rangle \nonumber \\
 &&\ge \frac14 \left\langle \tr (h^{-1}\dot h)^2 \right\rangle \ge 
 \frac{1}{4n} \left\langle \left[\tr \left(h^{-1}\dot h\right) \right]^2 \right\rangle\, , 
 \end{eqnarray}
 where the second inequality,  which is saturated when the $n\times n$ matrix $h^{-1}\dot h$ is proportional to the identity matrix, follows from the fact that $\tr[M-n^{-1}\tr M]^2 \ge0$ for any $n\times n$ matrix $M$; the resulting inequality is again a necessary condition for the SEC.
This last step 
 allows us to rewrite the right hand side of (\ref{halfway}) using the identity
 \begin{equation}
     \left[ \frac12\tr \left(h^{-1}\dot h\right) \right]^2   \equiv (d-2)^2 \left(\dot\Omega/\Omega\right)^2 +
     X^2 -2(d-2) X (\dot\Omega/\Omega) \, . 
 \end{equation}
 This yields the following integrated SEC inequality 
\begin{eqnarray}\label{keyineq}
&&- (d-1)\left[ (\ddot S/S) + (\dot S/S) \langle \dot\Omega/\Omega\rangle  \right]
 - \langle \ddot\Omega/\Omega\rangle + \left(1- a^{-2}\right)   \langle (\dot\Omega/\Omega)^2\rangle \nonumber \\
 && +\frac{(n-1)}{n}\left\langle X^2\right\rangle  + \left(1+\frac{2(d-2)}{n}\right) \left\langle X (\dot\Omega/\Omega) \right\rangle \ge 0\, , 
 \end{eqnarray}
 where we have introduced the constant
 \begin{equation}
     a= \sqrt{\frac{n}{(D-2)(d-2)}}\, . 
 \end{equation}
 This inequality  is saturated when the $D$-dimensional averaged SEC is saturated {\it and} when only the size of the compact space $B$ (and not its shape) is time-dependent.

If there is a time $t_0$ for which all first derivatives are zero then the integrated SEC  inequality (\ref{keyineq}) implies that,  at this time, 
 \begin{equation}\label{Ob}
- (d-1) (\ddot S/S)
 - \langle \ddot\Omega/\Omega\rangle \geq 0\ \qquad \qquad (t=t_0)\ , 
 \end{equation}
 which is the inequality found in \cite{Obied:2018sgi}. It does not exclude acceleration\footnote{
 Here we appear to have a disagreement with Section 2.4 of  \cite{Obied:2018sgi}, not only because \eqref{Ob} does not rule out transient 
acceleration (as explained in what follows), but also because the acceleration bound (2.12) of \cite{Obied:2018sgi} is derived for a late-time 
(power-law) attractor  solution for which \eqref{Ob} does not apply because first derivatives are not zero.} because a negative contribution to the left hand side from a positive $\ddot S$ can be overcome by a positive contribution from a negative $\ddot\Omega$. 
 Indeed, this  is precisely the mechanism for the transient acceleration found in the time-dependent compactifications mentioned in the Introduction: the acceleration occurs as the volume of the
 compact space passes through a minimum value, corresponding to a maximum of $\Omega(t)$. 
 
 The issue that we wish to address is whether the SEC permits {\it late-time} acceleration. At late-times we expect some scaling solution 
 of the Einstein equations for which $\dot\Omega$ and $\dot S$ are, generically, never zero, in which case  it is (\ref{keyineq}), rather than (\ref{Ob}), that has implications for
late-time behaviour.

 %%%%%%%%%%%%%%%%%%%%%
 \subsection{No-go theorems}
  
 If $\dot\Omega\equiv0$ then (\ref{keyineq}) reduces  to
 \begin{equation}\label{ddotS}
 \ddot S \le0\, , 
 \end{equation}
which tells us that the expansion of the universe is non-accelerating.  This result incorporates the GMN theorem but is more general for two reasons. Firstly, it applies 
to any FLRW spacetime, not just dS.  Secondly, $\dot\Omega\equiv0$ does not imply time-independence of the metric on $B$; it implies only (via the Einstein-frame condition) 
that the integral $\int_B \Omega^{d-2} {\rm vol}_B$ is constant (where ${\rm vol}_B$ is the volume $n$-form associated to the metric $h$).   

If $\dot\Omega\not\equiv0$ then there is no obvious conclusion to be drawn from the inequality (\ref{keyineq}), principally because the final term on the left hand side could be positive or negative. 
It could also be zero, and {\it is} zero if the  Einstein-frame condition is realized in the form 
\begin{equation}\label{unav}
\sqrt{\det h}\,  \Omega^{d-2} = \omega (y)  \, , \qquad \int_B d^n y \ \omega(y) = G_D/G_d\, . 
\end{equation}
In this case the first-order Einstein frame condition (\ref{EFCderived}) simplifies to 
\begin{equation}\label{ECsimple}
 \frac12\tr \left(h^{-1}\dot h\right) \equiv - (d-2)(\dot\Omega/\Omega) \qquad \left(\Leftrightarrow \ X\equiv 0\right). 
 \end{equation}
The inequality (\ref{keyineq}) then reduces to\footnote{This is equivalent to  the inequality found in \cite{Teo:2004hq} for $d=4$, but the subsequent analysis there limits  the results to vacuum solutions only.}
 \begin{equation}\label{simpineq}
 - (d-1)\left[ (\ddot S/S) + (\dot S/S) \langle \dot\Omega/\Omega\rangle  \right]
 - \langle \ddot\Omega/\Omega\rangle + \left(1- a^{-2} \right)   \langle (\dot\Omega/\Omega)^2\rangle \ge0\, . 
 \end{equation}
This generalizes the bound derived in \cite{Obied:2018sgi} to the  case where $\dot \Omega$ and $\dot S$ are different from zero. They are never zero 
for the accelerating scaling solutions that we shall exhibit later. 
 
 As before, the inequality (\ref{simpineq}) is saturated when the $D$-dimensional averaged SEC is saturated {\it and} only the size of the compact space $B$ is time-dependent.
It  excludes a late-time dS universe,  i.e. one for which $S(t) \sim  e^{Ht}$ for positive constant $H$.  To see this, observe that 
the  $\ddot S/S$ term will dominate (leading to a contradiction)  unless $\Omega$ also increases faster than 
 any polynomial, and  if this increase is faster than exponential  then the $\ddot\Omega/\Omega$ term 
 will dominate  (again leading to a contradiction), so we must suppose that $\Omega(y;t) = \tilde\Omega(y) e^{Jt}$ for constant $J$.
 The inequality (\ref{simpineq}) then yields the following quadratic inequality in $(H,J)$:
 \begin{equation}
 H^2 + HJ + \frac{(D-2)(d-2)}{n(d-1)} J^2 \le 0\, . 
 \end{equation}
This equation has no real solutions, so compactification to dS is excluded. 

This does not prove that the SEC forbids a time-dependent compactification to dS because the Einstein-frame condition (\ref{Eframe})  
could be satisfied even though its unaveraged form (\ref{unav}) is not, and in  this case we would have to return to the inequality
(\ref{keyineq}).   Nevertheless,  we think it worthwhile to explore the implications of the SEC given the assumption that the 
unaveraged Einstein-frame condition is satisfied. As far as we are aware, this unaveraged form has been assumed in all previous
work  on time-dependent cosmological compactifications. 

%%%%%%%%%%%%%%%%%%%
\subsection{Power-law acceleration}

Having confirmed that compactification to dS is excluded under the assumption that the Einstein-frame condition is satisfied in its
unaveraged form (\ref{unav}),  we now investigate whether this assumption allows a  time-dependent compactification to an FLRW universes 
undergoing late-time acceleration; i.e. we assume that $S \sim t^\eta$ at late time, for some constant $\eta$, and seek scaling solutions 
of this type.   In this case, the inequality (\ref{simpineq}) does not allow exponential growth of $\Omega$, so we shall suppose that  
$\Omega \sim t^\xi$ at late time, for some other constant $\xi$.  The asymptotic solution must then take the form\footnote{Strictly speaking, the SEC
also allows $S= t^\eta f_1(t)$,  and $\Omega=\Omega_0\ t^\xi f_2(t)$  if $t f'/f  \to 0 $ and $t^2f''/f  \to 0$ as $t \to \infty$, but this still  leads to the inequality
(\ref{powerlaw}).} 
 \begin{equation}\label{powers}
 S = t^\eta \, , \qquad \Omega = \Omega_0(y) t^\xi\, , 
 \end{equation}
 where a normalization constant for $S(t)$ is omitted because it can be absorbed into $\bar g_{ij}$. 
 The inequality (\ref{simpineq}) now becomes
 \begin{equation}\label{powerlaw}
\eta^2 + \eta\xi + \frac{(D-2)(d-2)}{n(d-1)} \xi^2 - \eta - \frac{1}{(d-1)} \xi \le 0\, .
\end{equation}
This defines the interior of an ellipse in the $(\eta,\xi)$ plane. For an accelerating universe we must also have $\eta>1$, but the line
$\eta=1$ intersects the ellipse so there is a region inside the ellipse for which $\eta>1$.  The largest
value of $\eta$ allowed by the SEC corresponds to the line $\eta=\eta_{\rm max}$ that is tangent to the ellipse, which occurs when
\begin{equation}
\eta_{\rm max}= \frac{ (d-1) \left(2-a^2\right) +2 \sqrt{(d-1)^2  -a^2 (d-1) ( d-2) }} {(d-1) \left(4-a^2 (d-1)\right)} <\frac43\ ,
\end{equation}
where the upper bound is approached as both $d\to \infty$ and $n\to \infty$; in contrast $\eta_{\rm max}\to 1$ as as $d\to\infty$ for fixed $n$. 
Within the region $1\le \eta \le \eta_{\rm max}$ the parameter $\xi$ is bounded by 
\begin{equation}\label{ximin}
- \frac{n}{D-2} \le \xi \le 0\, . 
\end{equation}

\begin{figure}[h!]
\centering
\includegraphics[width=0.6\textwidth]{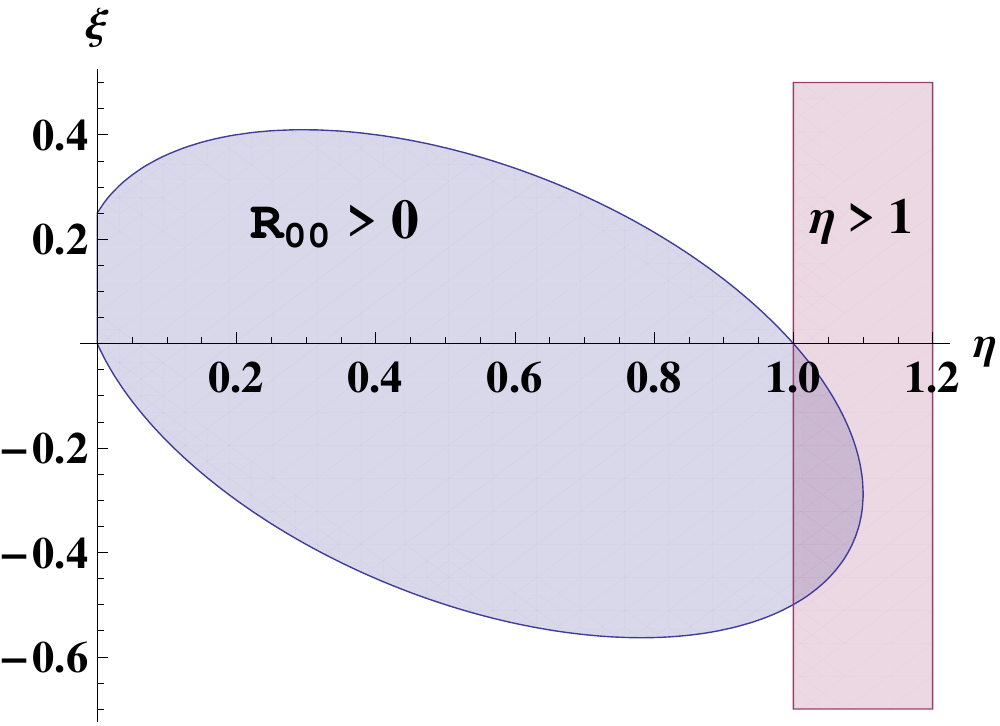}
\caption{The SEC  $R_{00}\geq 0$  is the region inside the solid ellipse (here $D=6,\ d=4$). The part of the region $\eta>1$ inside this ellipse describes accelerating universes satisfying SEC (this region is inside the  much larger  region allowed by DEC, which we determine in section 3.2). }
\label{nufa}
\end{figure}

The restriction to negative $\xi$ tells us that the compact space is expanding\footnote{If the bound  (\ref{Ob}) derived in \cite{Obied:2018sgi} were to be used then one would conclude (incorrectly) that 
$\xi$ is positive, and hence that the compact space is contracting, but this bound does not apply here since neither $\dot S$ nor $\dot\Omega$ is zero at late times.}. This implies an eventual decompactification, which we now investigate.

%%%%%%%%%%%%%%%%%%%%%%%%%%%%%%%%%%%
\subsection{Decompactification}

Let  $\varphi(t)$ be the scale factor for the $n$-dimensional compact space; i.e. 
\begin{equation}
\varphi^n  \propto \sqrt{\det h}\, . 
\end{equation}
Since 
\begin{equation}
\frac{d}{dt} \sqrt{\det h} = - (d-2) (\dot\Omega/\Omega) \sqrt{\det h}  = -\frac{(d-2) \xi}{t} \sqrt{\det h} \, , 
\end{equation}
we have
\begin{equation}
\dot\varphi = -\frac{(d-2)\xi}{nt} \varphi  \, , 
\end{equation}
which has the solution 
\begin{equation}\label{sig}
\varphi\  \propto \ t^\sigma\, , \qquad \sigma = - \frac{(d-2)\xi}{n} \, . 
\end{equation}
When expressed in terms of $\sigma$  the restriction  (\ref{ximin}) on $\xi$  implies (for $\xi\ne0$) that 
\begin{equation}
0 < \sigma \le \frac{d-2}{D-2} \, . 
\end{equation}
This confirms that the compact space is expanding  (as $\sigma>0$) and it additionally tells us that {\it this} expansion is decelerating (as $\sigma<1$). 

A feature of additional compact dimensions is that the harmonic expansion of fields on these dimensions will lead to an infinite sequence of  Kaluza-Klein  (KK) particles with some lowest 
mass set by the  scale of the compact space. Let $M_0$ be the mass of the lightest KK particle at time $t_0$; then the corresponding mass at a later time $t$ will be 
\begin{equation}
M(t) = \left(t_0/t\right)^{\sigma-\xi} M_0 \qquad (t>t_0). 
\end{equation}
This formula takes into account the the time-dependence arising from the warp factor $\Omega$ in addition to the the time-dependence of the volume of the compact space metric coming 
from its scale factor $\varphi$. 
From the formula for $\sigma$ in (\ref{sig}) we see that 
\begin{equation}
\sigma-\xi = -\frac{(D-2)}{n} \xi 
\end{equation}
and hence, since $\xi \le0$, 
\begin{equation}
M(t) =  \left(t_0/t\right)^{(D-2)|\xi|/n} M_0\, . 
\end{equation}
This is a decreasing function of time, and the existence of the KK tower of particles will have observable consequences at some ``decompactification time''  $T>t_0$ when $M(T)$ is of the order
of particle masses accessible to particle accelerators. The relation between  $T$ and $M(T)$ is
\begin{equation}
\frac{M_0}{M(T)} = \left(\frac{T}{t_0}\right)^{\frac{(D-2)|\xi|}{n}}\, ,  
\end{equation}
but we also have $S(T) = (T/t_0)^\eta$, and hence 
\begin{equation}
S(T) =  \left(\frac{M_0}{M(T)}\right)^\gamma \, ,  \qquad \gamma = \frac{n\eta}{(D-2)|\xi|} \ge 1\, ,
\end{equation}
where the lower bound on $\gamma $ is a consequence of the bound  (\ref{Pint}) on $\xi $ and the assumption that $\eta>1$.
This tells us how much the  FLRW universe has expanded prior to its effective decompactification to a $D$-dimensional universe.  Of course, this is an 
approximation because we are using only the late-time solution. 

To get an idea for the likely numbers in any attempt to apply these ideas to a realistic model, let us take 
$M_0$ to be the Planck mass and $M(T)$ the mass of the Higgs boson. 
Then $M_0/M(T) \sim 10^{17}$, so 
\begin{equation}
S(T) \sim 10^{17\gamma}\, . 
\end{equation}
This is arbitrarily large because there is no upper bound on $\gamma$; this means that $T$ could be much longer than the current age of the universe. 
However, one might suspect that fine tuning is needed for $\gamma \gg1$, so  it is of interest to ask how large $\gamma$ must be to avoid a conflict with the 
current lack of evidence for extra dimensions.  If $t_0$ is the time at the end of inflation (which is the earliest time that we could reasonably choose) then we 
would need $S(T) \gg 10^{30}$ to be sure of consistency with the current non-appearance of  KK particles.   This imposes a phenomenological bound that is
approximately $\gamma>2$. 

%%%%%%%%%%%%%%%%%%%%%%%%%%%%%%%%%%%%%%%%%%%
\subsection{SEC Redux}

Now that we have  found a possibility for compactification to an accelerating FLRW universe, we must recall 
that the integrated SEC inequality is merely a necessary condition for validity of the SEC. This means that we must return to the unintegrated condition $R_{00}\ge0$ and ask whether this is also satisfied. As we are now assuming that the Einstein-frame condition is satisfied in unintegrated form, we may use this to simplify the expression
(\ref{r00}) for $R_{00}$, and hence the unintegrated SEC $R_{00}\ge0$;  one finds that
\begin{eqnarray}\label{unint}
&&- (d-1)\left[ (\ddot S/S) + (\dot S/S)(\dot\Omega/\Omega) \right]
 - (\ddot\Omega/\Omega) + \left(1- a^{-2}\right)(\dot\Omega/\Omega)^2 \nonumber \\
 &&\  + \ \frac{1}{d} \Omega^{2-d}\nabla^2\Omega^d \ \ge 0\, . 
\end{eqnarray}
This reduces to (\ref{simpineq}) upon taking the average over $B$, but the unaveraged inequality 
involves an additional term proportional to $\Omega^{2-d}\nabla^2\Omega^d$. Since this term averages to zero
it must be negative somewhere on the compact space $B$ unless it is identically zero on $B$, which would require
$\Omega$ to be (time-dependent) constant on $B$. 

What this means is that compactification metrics with $\Omega$ of the form assumed above in (\ref{powers})
are not guaranteed to satisfy the SEC inequality, even if they do satisfy the integrated SEC inequality, unless
we further assume that $\Omega_0$ is constant on $B$.  It may be that for some non-constant choices of $\Omega_0$ the
unintegrated SEC inequality (\ref{unint}) will be satisfied, but more work is required to answer this question.
For this reason, we shall assume in the following section that $\Omega_0$ is a constant.

%%%%%%%%%%%%%%%%%%%%%%%%%%%%%%%%%
%%%%%%%%%%%%%%%%%%%%%%%%%%%%%%%%%%
\section{The late-time stress tensor}

We have now seen that the SEC permits compactification to eternally accelerating universes for which 
\begin{equation}\label{late}
S(t) = t^\eta\, , \qquad \Omega(t)=  \Omega_0\,  t^\xi \, ,  
\end{equation}
but we do not yet  know anything about the stress tensor that makes this possible. For example, we do not yet know  whether it could be zero, although we shall soon see that this is not a possibility. 

A calculation, using (\ref{late}) and (\ref{ECsimple}), yields
\begin{eqnarray}\label{Ricci1}
R_{00} &=& -\frac{(d-1)}{t^2} \left\{ \eta^2 +\eta\xi + \frac{(D-2)(d-2)}{n(d-1)} \xi^2 -\eta  - \frac{1}{(d-1)} \xi \right\}\ ,\nonumber\\
R_{ij} &=& \left\{ t^{2(\eta-1)} (\xi+\eta)\left[(d-1)\eta-1\right] + (d-2)k\right\}\bar g_{ij}\, .
\end{eqnarray}
The space-curvature term proportional to $k$ in the expression for $R_{ij}$ is non-leading as $t\to \infty$ when $\eta>1$, which is 
precisely the condition for accelerated expansion. For this reason we ignore this term in what follows, although we will re-instate
it later when we consider the $\eta=1$ case, since zero acceleration for $t=\infty$ does not exclude a positive acceleration for large
but finite $t$. 

We shall further suppose here that only the volume of $B$ in the metric $h$ is time-dependent; i.e. 
 \begin{equation}\label{tildeh}
 h_{\alpha\beta} = \varphi^2(t) \tilde h_{\alpha\beta}(y) \, , 
 \end{equation}
 where $\varphi$ is a scale factor for the compact space $B$.  In this case, 
 \begin{equation}
 \tr \left(h^{-1}\dot h\right) = 2n  (\dot\varphi/\varphi)\, , \qquad \tr (h^{-1}\dot h)^2 = 4n  (\dot\varphi/\varphi)^2\, .  
 \end{equation}
Using (\ref{ECsimple}) again we learn that 
 \begin{equation}
 \varphi\propto  t^{- (d-2)\xi/n}\, .
 \end{equation}
 In addition, we shall suppose that the metric $h$ on $B$ is Ricci  flat; we return later to consider the implications
 of allowing it to be a generic Einstein metric on $B$. Under this Ricci-flat assumption, a
 further calculation shows that 
\begin{equation}\label{Ricci2}
R_{\alpha\beta} =- t^{-2(1+\xi)} \left[(d-2)/n\right] \left[(d-1)\eta-1\right]\xi  \, h_{\alpha\beta} \, ,
\end{equation}
from which we deduce the following expression for the Ricci scalar:
\begin{equation}
R= t^{-2(1+\xi)} \left\{ d(d-1)\eta^2 + 2(d-1)\eta\xi + \frac{(d-2)(D-2)}{n}\xi^2 -2(d-1)\eta -2\xi\right\}\, . 
\end{equation}
All other  components of the Ricci tensor are zero. 

From these results we find that the non-zero components of the Einstein tensor take the form\footnote{We recall that $g_{ij}= S^2\bar g_{ij}$.}
\begin{eqnarray}\label{Ein}
G_{00}^{(D)} =  \frac{A}{t^2}\, , \qquad  G^{(D)}_{ij} = \frac{B}{t^2} \, g_{ij} \, , \qquad  G^{(D)}_{\alpha\beta} = \frac{C}{t^{2(1+\xi )}} \,   h_{\alpha\beta}\, , 
\end{eqnarray}
where the three constants $(A,B,C)$ are (after setting $k=0$ for the reasons explained above)
\begin{eqnarray}\label{ABC}
A&=& \frac{(d-2)}{2} \left[ (d-1)\eta^2 - \frac{(D-2)}{n} \xi^2 \right] \, , \nonumber \\
B&=&  \frac{(d-2)}{2} \left[- (d-1)\eta^2 - \frac{(D-2)}{n} \xi^2 + 2\eta\right] \, , \nonumber\\
C&=& - \frac{d(d-1)}{2} \eta^2 - \frac{(d-1)(D-2)}{n}\eta\xi - \frac{(d-2)(D-2)}{2n}\xi^2 \\
&& \qquad  + \ (d-1)\eta  + \frac{(D-2)}{n} \xi \, .  \nonumber 
\end{eqnarray}
The contracted Bianchi identities satisfied by the Einstein tensor are equivalent to the identity
\begin{equation}\label{linid}
\left[2+ \xi -(d-1)\eta\right]A -(d-1)(\eta+\xi)B + (d-2)\xi C  \equiv 0\, , 
\end{equation}
as substitution for $(A,B,C)$ in terms of $(\eta,\xi)$ confirms. 

The {\it vacuum} Einstein equations in $D$-dimensions now reduce to $A=B=C=0$, but elimination of $\xi$ from the 
two equations $A=B=0$ yields the equation $\eta[(d-1)\eta-1]=0$, which implies that $\eta<1$; 
this agrees with the result of  \cite{Teo:2004hq}, where the vacuum equations were used to simplify the SEC inequality.

We now know that the stress
tensor required for $\eta>1$ must be non-zero. We shall take its non-zero components to be 
\begin{equation}\label{Dstress}
T_{00} = t^{-2} \rho_0\, , \qquad T_{ij} = t^{-2} P_0 g_{ij}\, , \qquad T_{\alpha\beta} = t^{-2(1+\xi)}  P^{({\rm int})}_0 h_{\alpha\beta}\, , 
\end{equation} 
for constants $\{\rho_0, P_0, P^{({\rm int})}_0\}$.  This choice is consistent with the Bianchi identities provided that
\begin{equation}\label{linearR}
\left[2+ \xi - (d-1)\eta\right]\rho_0 - (d-1)(\eta+\xi) P_0 + (d-2)\xi P^{({\rm int})}_0 =0\, . 
\end{equation}
For a convenient choice of units, the late time Einstein equations are now
\begin{equation}\label{ABC-Einstein}
A= \rho_0\, \qquad B= P_0\, , \qquad C= P^{({\rm int})}_0\, ,  
\end{equation}
but only two of these three equations are independent because of the linear relations (\ref{linid}) and (\ref{linearR}).  We may take the first and third
equations as the independent ones if $\xi=0$, but it is then straightforward to show that  the SEC cannot be satisfied for $\eta>1$;  this is expected because $\xi=0$
implies a time-independent metric for the compact manifold $B$.   We may therefore assume that $\xi\ne0$, and in this case we may take the first two of the equations (\ref{ABC}) as 
the two independent equations; these are 
\begin{eqnarray}\label{density}
\rho_0 &=& \frac{(d-2)}{2}\left[(d-1)\eta^2 - \frac{(D-2)}{n}\xi^2\right] \, , \nonumber \\
P_0 &=& -\frac{(d-2)}{2}\left[(d-1)\eta^2 -2\eta + \frac{(D-2)}{n} \xi^2\right] \, . 
\end{eqnarray}
Using $\eta>1$ and $|\xi|<n/(D-2)$,  from (\ref{ximin}), one may easily show that 
\begin{equation}
\rho_0 > 0 \, , \qquad P_0 <0\, . 
\end{equation}

The pressure constant $P^{({\rm int})}_0$ is determined
by the linear relation (\ref{linearR}) (or, equivalently, by the third Einstein equation):
\begin{equation}\label{Pint}
P^{({\rm int})}_0 = -\frac{(d-1)}{2}\left[d\eta^2-2\eta\right] - \frac{(D-2)}{n}\xi\left[(d-1)\eta -1 +  \frac{(d-2)}{2}\xi\right]\, . 
\end{equation}
Using again the restriction (\ref{ximin}) on the range of $\xi$ allowed by the SEC, and the assumption that $\eta>1$, we have
\begin{equation}
P^{({\rm int})}_0  < - \frac{(d-2)(d-3)}{2} <0\, . 
\end{equation}

The stress tensor defined by
(\ref{Dstress}) suggests some anisotropic generalization of
a perfect fluid (possibly along the lines of 
of \cite{Borlaf:1998qm}) with the time-dependence determined  in terms of the two scale factors $(S,\varphi)$ by some analog of the 
continuity equation. If so, one might expect some field theoretic realization, possibly with the fields of $D=10/11$ supergravity. 
This is an interesting open problem  but the effective $d$-dimensional stress tensor is a perfect fluid with a scalar field theoretic realization, 
as we now discuss. 

%%%%%%%%%%%%%%%%%%%%%%%%%%%%%
\subsection{A new no-go theorem}

We have considered a class of cosmological compactification solutions of the $D$-dimensional Einstein field equations 
for which the  Einstein-frame metric in the lower dimension is an FLRW spacetime with scale factor $S^\eta$, and we have shown that 
the SEC in $D$ dimensions is compatible with $\eta>1$, i.e. accelerated expansion.  The Einstein tensor for this FLRW metric has 
non-zero components 
\begin{eqnarray}\label{Ein}
G_{00} = \frac{A_0}{t^2} \equiv  \rho \ , \qquad  G_{ij} = \frac{B_0}{t^2} g_{ij} \equiv P\, g_{ij} \, , 
\end{eqnarray}
where
\begin{equation}
A_0 =  \frac{(d-2)}{2} \left[(d-1)\eta^2 \right] \, , \qquad B_0 =  -\frac{(d-2)}{2} \left[ (d-1)\eta^2  - 2\eta\right] \, . 
\end{equation}
These are the coefficients $A$ and $B$ of (\ref{ABC}) at $\xi=0$, which are those required for consistency with the Bianchi identity satisfied by the Einstein tensor for the Einstein-frame $d$-metric. 

The $d$-dimensional stress tensor is that of a perfect fluid with mass density $\rho$ and pressure $P$ satisfying the continuity equation 
\begin{equation}
\dot \rho = -(d-1)\left(\rho+P\right) (\dot S/S) \, , 
\end{equation}
as one may verify.  The equation of state is 
\begin{equation}
P = w\rho \, , \qquad w = B_0/A_0 = -1 + \frac{2}{(d-1)\eta}  \, . 
\end{equation}
This equation of state is what one finds, under the assumptions of homogeneity and isotropy, for a scalar field in $d$-dimensions 
with a positive exponential potential, which has a subcritical exponent when $\eta>1$.  The only restriction on $w$ implied by this scalar field realization 
is that it must lie in the interval $[-1,1]$. However, we  have found, by requiring this spacetime to arise from a cosmological compactification  
of  a higher-dimensional gravitational theory with stress tensor satisfying the SEC, that 
\begin{equation}
\eta < \eta_{\rm max} <4/3\, , 
\end{equation}
where the second inequality, which is independent of both $d$ and $D$,  translates to the following lower bound on $w$:
\begin{equation}
w > -1 + \frac{3}{2(d-1)}\, . 
\end{equation}

In subsection 2.2 we concluded (based on assumptions stated there) that a dS universe cannot be obtained by compactification even if the compact space and 
warp factor are allowed to be time dependent. This conclusion (along with the assumptions on which it depends) is in agreement with earlier work on a
generalization of the GMN no-go theorem applying to time-independent compactifications. In the current context, this generalized no-go theorem could be 
rephrased as the statement that $w=-1$ is excluded, but we have now arrived (under the {\it same} assumptions) at a  much stronger  restriction on $w$; for $d=4$ this restriction is the lower bound $w > -1/2$.  Of course, this statement applies to a perfect fluid with an equation of state $P=w\rho$ for constant $w$, which is (unless $w=-1$)
equivalent to the assumption of power-law expansion, which we can only expect to be valid at late times.

%%%%%%%%%%%%%%%%%%%%%%%%%%%
\subsection{The Dominant Energy Condition}

The SEC on the $D$-dimensional stress tensor is 
\begin{equation}\label{SEC2}
(D-3) \rho_0 + (d-1)P_0 + n P^{({\rm int})}_0 \ge 0\, . 
\end{equation}
As a check one may use the Einstein equations in the form (\ref{ABC-Einstein}) and the expressions of (\ref{ABC}) for $(A,B,C)$ to show that 
this implies the inequality (\ref{powerlaw}).  As we have seen, this inequality is compatible with acceleration, i.e. with $\eta>1$, but for any 
such choice of $(\eta,\xi)$ we need to check that the DEC is satisfied. This condition states that  no component of the stress tensor may have a magnitude greater than the energy density, which must  be non-negative. In the present context, this is equivalent to 
\begin{equation}
\rho_0 \ge 0 \, , \qquad \rho_0 \pm P_0 \ge 0\, , \qquad \rho_0 \pm P^{({\rm int})}_0 \ge 0 \, , 
\end{equation}
for any choice of the signs. As we already know that $\rho_0>0$ and that both $P_0$ and $P^{({\rm int})}_0$ are negative when the SEC is satisfied, 
we only need to check positivity of $\rho_0+P_0$ and $\rho_0+ P^{({\rm int})}_0$. We consider them in turn: 

\begin{itemize}

\item $\rho_0 + P_0$. In this case we have from (\ref{density}) that 
\begin{eqnarray}
\rho_0 + P_0  &=& (d-2)\left[\eta - \frac{(D-2)}{n} \xi^2\right]\nonumber \\
&>& (d-2)\left[ 1- \frac{n}{(D-2)}\right] = \frac{(d-2)^2}{D-2} >0\, , 
\end{eqnarray}
where the inequality follows from $\eta>1$ and $|\xi|< n/(D-2)$. 

\item $\rho_0 + P^{({\rm int})}_0$. In this case it is simplest to solve the linear relation (\ref{linearR}) for $P_0$ in terms of $\rho_0$ and 
$P^{({\rm int})}_0$, and  then substitute the result into the SEC condition (\ref{SEC2}). This yields the following inequality:
\begin{equation}
\left[n\eta + (D-2)\xi\right] \left(\rho_0+ P^{({\rm int})}_0\right) \ge 2(\eta-1)\rho_0 \, . 
\end{equation}
The right hand side is positve for $\eta>1$. The coefficient $[n\eta + (D-2)\xi]$ is also positive for $\eta>1$ as a consequence of the bounds
on $\xi$ imposed by the SEC. We thus conclude that 
\begin{equation}
\rho_0 + P^{({\rm int})}_0 >0\, , 
\end{equation}
when $\eta>1$, as a consequence of the SEC.

\end{itemize}
This concludes our demonstration, for the specific power-law type cosmological compactification under consideration,  that the DEC is a consequence of the 
SEC when the lower-dimensional FLRW universe is undergoing accelerated expansion.  Taken together with our finding that the SEC 
conditions are compatible with acceleration, we conclude that the combined SEC and DEC conditions in the higher dimension do not exclude
the possibility of compactification to an eternally accelerating FLRW universe (to which other solutions could asymptote at late time).

%%%%%%%%%%%%%%%%%%%%%%%
\subsection{Asymptotic zero acceleration}

As mentioned earlier, the $\eta=1$ case of the power-law solutions considered above should be taken into account in any discussion of late-time acceleration
because a universe expanding with zero acceleration may be approached asymptotically by one that has non-zero positive acceleration at late times. 
Allowing for $\eta=1$ in the above analysis so far does not change the conclusions, as long as $k=0$; i.e. as long as the $(d-1)$-space has zero curvature.
However, setting $\eta=1$ in the Ricci tensor expressions of (\ref{Ricci1}) and (\ref{Ricci2}) yields 
\begin{eqnarray}
R_{00} &=& -\frac{(d-2)}{t^2} \xi \left[1+ \frac{(D-2)}{n} \xi \right]\, , \nonumber \\
R_{ij} &=&  (1+ k + \xi)(d-2) \bar g_{ij} \, , \nonumber \\
R_{\alpha\beta} &=& - \frac{(d-2)^2}{n} t^{-2(1+\xi)}\xi\,  h_{\alpha\beta} \, .  
\end{eqnarray}
The space curvature $k$ is now relevant. 

We consider here only the vacuum Einstein equations. These have a solution for  $k=-1$, provided that $\xi=0$, which implies that the metric on the compact space is  time-independent; this is the 
compactification to a Milne universe discussed in \cite{Gibbons:1986xp}. In general, one may expect there to exist
initial conditions that lead  to accelerating universes  that are asymptotic, at late time,  to a zero-acceleration universe; a special feature of this case is that, despite the late time acceleration,  there is no future cosmological event horizon \cite{Boya:2002mv}. Examples were found in  \cite{Jarv:2004uk} for cosmological  models with ``double exponential potentials'' (for which exact solutions may also be found \cite{Kan:2018mha}).

%%%%%%%%%%%%%%%%%%%%%%%%%%
\subsection{Generic Einstein metric for the compact space}

So far we supposed the metric $h$ on the compact space $B$ to be Ricci-flat. More generally, we could 
suppose that it is an Einstein metric, in which case $\tilde R_{\alpha\beta} = (n-1)K\, \tilde h_{\alpha\beta}$ for constant $K$. In this case, 
the expression of (\ref{Ricci2}) is replaced by 
\begin{equation}\label{Ricci3}
R_{\alpha\beta} =- t^{-2(1+\xi)}\left\{  \left[(d-2)/n\right] \left[(d-1)\eta-1\right]\xi  - t^{2\left[ 1+ \frac{(D-2)}{n}\xi\right]} (n-1) K \right\}\, h_{\alpha\beta} \, . 
\end{equation}
Let us define 
\begin{equation}
\xi_0 = - \frac{n}{(D-2)}
\end{equation}
and consider in turn the implications of choosing $\xi$ to be less than, greater than or equal to $\xi_0$:
\begin{itemize}

\item $\xi< \xi_0$. The curvature term proportional to $K$ is subleading at late times and can be ignored, so
we are back to the $K=0$ case.

\item $\xi>\xi_0$.  The curvature term proportional to $K$  dominates and there is no vacuum solution of the assumed 
power-law form. 

\item $\xi=\xi_0$. In this case the SEC, which is unaffected by the curvature of space, requires 
\begin{equation}
\eta^2 - \left(1+ \frac{n}{(D-2)} \right)\eta + \frac{n}{(D-2)}  \le 0\,  \quad \Rightarrow \quad |\xi_0| \le  \eta \le 1\,  . 
\end{equation}
Only $\eta=1$ (which implies saturation of the SEC bound) is relevant to the possibility of late-time acceleration, 
and in this case we should also allow for non-zero $k$.  Again restricting to vacuum solutions (i.e. all components of the Ricci tensor are zero), we find that 
there is just one of them,  provided that we choose (recall that $k$ is un-normalized)
\begin{equation}
k= - \frac{(d-2)}{(D-2)}\, , \qquad K= - \frac{(d-2)^2}{(D-2)(n-1)} \, . 
\end{equation}
\end{itemize}
We remark that  the SEC is also saturated for $\xi=\xi_0$ when $\eta= n/(D-2)$. In this case, $\eta <1 $ and there is a vacuum solution for $K=-(d-2)^2/(D-2)^2$ provided
that $k=0$. This decelerating scaling solution is the late-time limit of the hyperbolic compactification exhibiting transient cosmic acceleration 
found in \cite{Townsend:2003fx}.

%%%%%%%%%%%%%%%%%%%%%%%%%
%%%%%%%%%%%%%%%%%%%%%%%
\section{Summary and discussion}

We have explored the constraints on cosmological compactifications implied by the strong energy condition (SEC) on the higher-dimensional stress tensor.
It is known that this condition rules out compactification to de Sitter space  (dS) if the compact space is non-singular and time-independent. This no-go theorem follows 
from a simple inequality implied by the SEC on the scale factor of de Sitter space viewed as an FLRW universe. A new result of this paper is the corresponding inequality \eqref{keyineq}  
for compactification to any Einstein-frame FLRW universe, allowing for any time dependence compatible with generic FLRW isometries. However, this inequality does not lead directly 
to any interesting conclusions -- in particular, it does {\it not} obviously rule out compactifications to de Sitter.  This is mainly because the Einstein-frame condition   involves an averaging  over the 
compact space. In known examples  the Einstein-frame condition is satisfied in an unaveraged form, and if this is assumed the SEC inequality simplifies 
dramatically to the form \eqref{simpineq};   the inequality used in  \cite{Obied:2018sgi} is then found by ignoring terms involving first derivatives, but their inclusion is 
essential to our analysis of the implications of the SEC for late-time behaviour. 

Many examples are known of time-dependent compactifications of D=10/11 supergravity theories for which there is a period of acceleration, in Einstein frame, but the late time universe in all these 
examples is decelerating. Since these supergravity theories satisfy the SEC, this might be a consequence of the SEC. However, we have shown, by example, 
that the SEC, by itself, does {\it not} prevent compactification to (Einstein frame)  universes that undergo late-time accelerated expansion. Our examples 
are scaling solutions that represent a time-dependent compactification to an FLRW universe that undergoes an {\it eternal} accelerated expansion (and hence, in particular, at late times) 
but we expect these to act as late-time attractor solutions for more general FLRW compactifications. To verify this we would first need a better understanding  of the $D$-dimensional 
stress tensor that makes such compactifications possible;  at present  we know only its late-time form, which suggests some anisotropic generalization of 
a perfect fluid. 

What we {\it can} say of our examples is that they are compactifications to FLRW universes filled with a perfect fluid that has a constant pressure to energy density ratio $w$, on which 
the higher dimensional SEC imposes a lower bound, which is $w>-1/2$ for $d=4$. The constancy of $w$ is implied by the power-law expansion, which is accelerated expansion
when (for $d=4$) $w<-1/3$. This shows, in particular, that the SEC can be violated in the lower dimension even though it is not violated in the higher dimension. We should stress
that the lower bound on $w$ (which excludes the dS case of $w=-1$) is a {\it late-time} result since we can expect $w$ to be constant only at late times (and then only if $w<-1/3$). 
There is therefore no conflict with the fact that observations are consistent with a present value of $w=-1$, especially as they are also consistent with increasing $w$  \cite{Aghanim:2018eyx}.

Another feature of our examples is that the compact space is also expanding, which implies a {\it late-time decompactification}; in practice this could be defined to occur at a 
decompactification time $T$ at which the effects of massive Kaluza-Klein (KK) particles become apparent in accelerator experiments. There could then be a conflict with observations 
if $T$ is less than the current age of our Universe. However, a remarkable property of our examples is that {\it accelerated expansion of the FLRW universe requires a decelerating 
expansion of the compact space}. Assuming that the lightest KK particle would become observable when its mass decreases to the Higgs mass, we have shown that the
decompactification time is arbitrarily large and, generically, much longer than the current age of our Universe.

%%%%%%%%%%%%%%%%%%%%%%%%%
\subsection{No-acceleration bounds on exponential scalar potentials}

An alternative response to our results would be to ask why we should have expected to extract so much from the SEC, which has no obvious 
fundamental significance. Quite possibly, our examples of  cosmic late-time acceleration from compactification cannot be realized in string/M-theory 
for other reasons. Or it could be ruled out by other general considerations, such as those mentioned in the introduction. As also mentioned there, 
the proposed cosmological swampland bound  of \cite{Obied:2018sgi} is of the type that can be easily deduced for positive exponential potentials  
in one-scalar theories if one supposes that late-time acceleration is impossible.  As promised in the Introduction, we conclude  with a comment on this bound. 

For a one-scalar theory with a positive exponential potential $V$, there will be cosmological solutions that undergo accelerated expansion at late time unless 
$|\nabla V|/V >c$, with $c = {\cal O}(1)$, which is the swampland bound proposed in \cite{Obied:2018sgi}.  However,  the situation is very different for a  
two-scalar model,  even for a positive exponential potential that is a function of  only one of the two fields;  in this case the no-acceleration bound on the 
coefficient $V'/V$  can depend on the interaction between the two scalar fields.

An example is provided by the following Lagrangian for  a dilaton field  $\sigma$ and an axion field $\chi$ whose vacuum values parametrize the hyperbolic space $H_2\cong Sl(2;R)/SO(2)$: 
\begin{equation}
L= R - \frac12\left[(\partial \sigma)^2 + e^{-\mu\sigma} (\partial\chi)^2\right] - \Lambda e^{-\lambda\sigma}\, . 
\end{equation}
We may assume that $\lambda>0$ but then $\mu$ could  be positive or negative; its  absolute value is inversely proportional to the radius of curvature of the $H_2$ target space. 
We also assume that $\Lambda>0$, since accelerating cosmologies are otherwise not possible. 
 For constant $\chi$ there is a scaling solution of the equations for  expanding FLRW cosmologies and the expansion is accelerating if  $\lambda<\lambda_c$ where $\lambda_c$ 
is a  `critical' value. For a $d$-dimensional spacetime, and in our conventions\footnote{Given in detail in \cite{Sonner:2006yn} except that here we have flipped the sign of $\mu$.}, 
\begin{equation}
\lambda_c = \sqrt{\frac{2}{d-2}}\, , 
\end{equation}
so that $\lambda_c=1$  for $d=4$.  

However, there is also a scaling scaling solution  with non-constant $\chi$ if $\mu>0$ and in this case the scale factor (in terms of standard FLRW time $t$) 
is \cite{Sonner:2006yn} 
\begin{equation}
S \propto t^\eta \, , \qquad \eta= \frac{\lambda + \mu}{(d-1)\lambda}\, . 
\end{equation}
Accelerated expansion occurs when $\eta>1$, which is equivalent to 
\begin{equation}
\lambda < \frac{\mu}{d-2}\, . 
\end{equation}
The right hand side is a new critical value for $\lambda$, which can be  larger than $\lambda_c$ for sufficiently large $\mu$. 

To avoid a possible 
late-time acceleration in this dilaton-axion model we must impose both  $\lambda>\lambda_c$  {\it and} $\mu< (d-2)\lambda$; given $\lambda>\lambda_c$, a sufficient condition on $\mu$ is
\begin{equation}
\mu <\mu_c = \sqrt{2(d-2)} \, , 
\end{equation}
so $\mu_c=2$ for $d=4$. This is a weak-coupling requirement. The stronger the coupling the more damped is the rolling down the exponential potential due to 
transfer of potential energy to the axion, which raises the critical value of $V'/V$ corresponding to a crossover from deceleration to acceleration.   The physics is analogous to the rolling of a disc 
of large moment of inertia down a hill under the influence of gravity; potential energy is transferred not only to kinetic energy of downward motion but
also to rotational kinetic energy, which slows the downward motion.

%%%%%%%%%%%%%%%%%%%%%%%%%
%%%%%%%%%%%%%%%%%%%%%%%%%

\section*{Acknowledgments}

We ae grateful to Thomas Van Riet for helpful correspondence.  JGR acknowledges financial support from projects 2017-SGR-929, MINECO
grant FPA2016-76005-C.
The work of PKT has been partially supported by STFC consolidated grant ST/L000385/1.

\setcounter{section}{0}

\end{document}